\documentclass{article}
\usepackage{amsmath, amssymb, amsthm, amsfonts, graphicx}
\newcommand{\pdt}[2]{\frac{\partial{#1}}{\partial{#2}}}
\newcommand{\spdt}[2]{\frac{\partial^2{#1}}{\partial{#2}^2}}
\DeclareMathOperator\grad{\text{grad}}

\begin{document}
\title{Can Sound Waves be Slowed?}
\author{Amey Joshi}
\date{14-Jul-2018}
\maketitle

\begin{abstract}
Dielectric fluids experience a striction force in the presence of an external electric field. Although
the striction force on the entire body of the fluid is usually zero, it does contribute to its 
deformations. In this paper, I show that the speed of sound in dielectric fluids is reduced when they 
are exposed to uniform electric fields of sufficiently high magnitude. Ferrofluids also experience a
similar striction force in presence of a magnetizing field. However, their low relative permeability
makes the effect very small.
\end{abstract}

\section{The mechanism of propagation of sound}
Sound waves are propagated by periodic compression and rarefaction of a material medium. Consider an
element of a material medium which is small at a macroscopic scale but big enough to have a large
number of molecules in it. A compression of such an element results in internal stresses within it 
which cause it to relax. The relaxation happens with such a magnitude that in the process the molecules 
in the element go farther from each other than they were at the equilibrium condition, leading to a 
state of rarefaction. While doing so it transfers its energy to the neighbouring element causing a 
compression in it. The overshooting from the condition of equilibrium to the other extreme of 
rarefaction again sets up internal stresses leading to a compression and the cycle continues forming 
a sound wave. This simplistic picture ignores several factors like the internal friction causing an 
eventual attenuation of the wave. Yet, it captures the key role played by internal stresses in the 
sustenance of sound waves.

\begin{figure}
\includegraphics{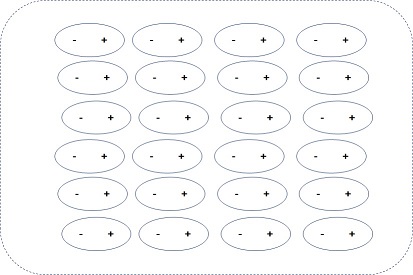}
\caption{Schematic diagram of a polarized dielectric}\label{f1}
\end{figure}

Let us continue with the same picture but now assume that the medium is a perfect dielectric exposed to 
a uniform external electric field. The field polarizes the molecules of the medium. Every element of 
the fluid now behaves like an electric dipole. If we also assume that the medium is a linear 
dielectric, as is the case with fluids, then the polarization is in the direction of the external 
electric field. Figure \ref{f1} shows a schematic diagram of a polarized dielectric. The elements of
the dielectric have their polarization lined along the external electric field. The picture depicts 
the polarization of elements of the medium and not the molecular dipole moments.
 
When a sound wave propagates through the medium, the polarized dipoles get closer to each other. 
Although the internal stresses will cause the them to go away from each other, the mild electrical
attraction between them delays their relaxation to a state of rarefaction. We therefore expect 
that the sound wave will propagate at a speed lower than it would in the absence of the electric 
field. 

In the following section we derive the equation of sound waves in a polarized dielectric fluid. We
are led to an expression for the speed of the wave in terms of partial derivative of the fluid's
relative permittivity with respect to its density. In order to calculate it, we need a relation between
the two quantities. Getting one is the goal of section 3. It also proves that the speed of sound in 
a polarized dielectric is always less than the speed of sound when there is no external electric
field. In section 4 we estimate the drop in speed for several dielectric fluids. Section 5 carries 
over the analysis of the dielectric fluids to ferrofluids. 

\section{Wave equation in the presence of electric field}
The motion of a dielectric fluid in the presence of an electric field is described by 
\begin{equation}\label{e1}
\rho\left(\pdt{\vec{u}}{t} + \vec{u}\cdot\grad{\vec{u}}\right) = 
-\grad{p} + \mu\nabla^2\vec{u} + \vec{f},
\end{equation}
where $\rho$ is the mass density, $\vec{u}$ the flow velocity, $p$ the pressure, $\mu$ the viscosity
and $\vec{f}$ the density of external body force. The velocity of the fluid is 
very small as compared to the speed of sound so that we can ignore the non-linear term on the
left hand side of equation \eqref{e1}. Further, we consider the sound waves to be generated in a
fluid which is otherwise at equilibrium. Therefore, we also ignore the viscous term on the right
hand side of equation \eqref{e1} so that we are left with
\begin{equation}\label{e2}
\rho\pdt{\vec{u}}{t} = -\grad{p} + \vec{f}.
\end{equation}
The body force density due to the external electric field is \cite{joshi2013stress}
\begin{equation}\label{e3}
\vec{f} = -\epsilon_0\frac{E^2}{2}\grad{\kappa} + 
\frac{\epsilon_0}{2}\grad{\left(E^2\rho\pdt{\kappa}{\rho}\right)},
\end{equation}
where $E = |\vec{E}|$ is the magnitude of the external electric field, $\kappa$ is the relative
permittivity of the material and $\epsilon_0$ is the permittivity of free space. We note that the
body force density does not depend on the direction of $\vec{E}$. When the electric
field is uniform, we can simplify equation \eqref{e2} as
\begin{equation}\label{e4}
\rho\pdt{\vec{u}}{t} = -\grad{p}_m,
\end{equation}
where $p_m$ is a modified pressure given by 
\begin{equation}\label{e5}
p_m = p + \epsilon_0\frac{E^2}{2}\left(\kappa - \rho\pdt{\kappa}{\rho}\right).
\end{equation}
Equation \eqref{e4} is one of the basic equations from which Lighthill \cite{lighthill2001waves} 
derives the equation of the sound waves in fluids, except that the hydrostatic pressure $p$ in his 
equation is replaced with the modified pressure $p_m$ in ours. The other basic equation in Lighthill's
treatment is that of conservation of mass. Following his treatment we get equation for the velocity 
potential $\phi$ as,
\begin{equation}\label{e6}
\spdt{\phi}{t} = c^2\nabla^2\phi,
\end{equation}
where $c$ is the speed of the sound wave given by
\begin{equation}\label{e7}
c^2 = \pdt{p_m}{\rho}.
\end{equation}
Using equation \eqref{e5} in the above equation we immediately get
\begin{equation}\label{e8}
c^2 = c_0^2 - \rho\epsilon_0\frac{E^2}{2}\frac{\partial^2\kappa}{\partial\rho^2},
\end{equation}
where $c_0 = \sqrt{\partial p/\partial\rho}$ is the speed of sound in the absence of electric field. In 
order to proceed further we need a relation between the relative permittivity and the mass density.
We get it in the following section.

\section{Clausius-Mossotti relation}
The dipole moment, $\vec{p}$, of a molecule in the presence of an external electric field, $\vec{E}$,
is given by $\vec{p} = \alpha\vec{E}$, where the constant $\alpha$ is called the molecular 
polarizability. The polarization of a fluid element of the kind described in section 1 is $\vec{P} = N
\langle\vec{p}\rangle$, where $N$ is the number of molecules per unit volume and $\langle\cdot\rangle$ 
denotes an average over the element. We shall find it convenient to write this as
\begin{equation}\label{e9}
\vec{P} = \rho\alpha_m\vec{E},
\end{equation}
where $\rho$ is the mass density and $\alpha_m = \alpha/m$, $m$ being the molecular weight. The
molecular polarizability is related to the relative permittivity (also called the dielectric
constant), $\kappa$ as \cite{reitz2008foundations}
\begin{equation}\label{e10}
\frac{\kappa - 1}{\kappa + 2} = \frac{\rho\alpha_m}{3\epsilon_0}
\end{equation}
Equation \eqref{e10} is called the Clausius-Mossitti relation, another form of which is
\begin{equation}\label{e10a}
\rho\pdt{\kappa}{\rho} = \frac{(\kappa - 1)(\kappa + 2)}{3}.
\end{equation}
Using equation \eqref{e10a} in \eqref{e8} we get
\begin{equation}\label{e11}
c^2 = c_0^2 - \frac{\epsilon_0 E^2}{9\rho}(\kappa + 2)(\kappa - 1)^2
\end{equation}
As the relative permittivity is always positive, the above equation confirms that the speed of
sound in a polarized dielectric medium is lower than it is in an unpolarized medium. For a fixed field 
strength, the decrease in speed of sound waves depends on the quantity
\begin{equation}\label{e12}
K = \frac{(\kappa + 2)(\kappa - 1)^2}{\rho},
\end{equation}
which depends solely on the properties of the dielectric fluid. In the next section we shall 
estimate the drop in speed of sound in a few dielectric fluids.

\section{Estimates of sound speed in dielectric fluids}
Speed of sound in absence of an electric field is given by
\begin{equation}\label{e13}
c_0^2 = \frac{B}{\rho},
\end{equation}
where $B$ is the bulk modulus of the fluid. It is not easy to get the bulk moduli of all the
dielectric fluids of interest. We therefore assume that they all have the same bulk modulus. A quick
perusal of the values of bulk moduli of the few liquids in the CRC Handbook of Physics and Chemistry
\cite{haynes2014crc} suggests that this is not a very bad assumption. We use this assumption to
estimate the $c_0$ for fluids other than water using the relation
\begin{equation}\label{e14}
\frac{c_{0l}}{c_{0w}} = \sqrt{\frac{\rho_w}{\rho_l}},
\end{equation}
where $c_{0l}$ and $c_{0w}$ are the speeds of sound in a dielectric liquid and water, whose densities
are $\rho_l$ and $\rho_w$ respectively. Table \ref{t1} summarizes the speed of sound in several
dielectric fluids exposed to a constant, uniform external electric field of $5$ kV/mm. This field
is lower than the dielectric breakdown of all liquids listed in the CRC Handbook of Physics and
Chemistry \cite{haynes2014crc} and we assume that none of the fluids listed below suffer a breakdown
at its value. All values in the table are in SI units and the second column has the temperature in 
kelvin at which the relative permittivity was measured. We assume that the density values available
in the CRC Handbook of Physics and Chemistry \cite{haynes2014crc} are appropriate for these 
temperatures.
\begin{table}[]
\begin{tabular}{lllllll}
\hline
\multicolumn{1}{c}{Dielectric} & \multicolumn{1}{c}{T} & \multicolumn{1}{c}{$\kappa$} & \multicolumn{1}{c}{$\rho$} & \multicolumn{1}{c}{$c_0$} & \multicolumn{1}{c}{$c$} & \multicolumn{1}{c}{\% slowdown} \\
\hline
Hydrogen fluoride       & 273.2 & 83.6     & 818    & 1636  & 1630.475 & 0.33\%      \\
Water                   & 293.2 & 80.1     & 998    & 1481  & 1476.720 & 0.29\%      \\
Formamide               & 293.2 & 111      & 1334   & 1281  & 1271.103 & 0.77\%      \\
Hydrogen cyanide        & 293.2 & 114.9    & 687.6  & 1784  & 1768.970 & 0.86\%      \\
Ethylene carbonate      & 313.2 & 89.78    & 1321   & 1287  & 1282.026 & 0.41\%      \\
N-methyl formamide      & 293.2 & 189      & 1011   & 1471  & 1414.544 & 3.87\%      \\
N-methyl acetamide      & 303.2 & 179      & 937.1  & 1528  & 1478.307 & 3.28\%      \\
N-acetyl methanolamine  & 298.2 & 96.6     & 1107.9 & 1406  & 1398.493 & 0.51\%      \\
2-Pyridine carbonitrile & 303.2 & 93.77    & 1081   & 1423  & 1416.404 & 0.46\%      \\
p-Nitroaniline          & 428   & 78.5     & 1219.2 & 1340  & 1336.288 & 0.27\%      \\
N-propyl acetamide      & 293.2 & 104      & 896    & 1563  & 1553.121 & 0.63\%      \\
N-propyl propinamide    & 298.2 & 118.1    & 898.5  & 1561  & 1546.342 & 0.93\%      \\
\hline    
\end{tabular}
\caption{Speed of sound in liquid dielectric exposed to uniform electric field of $5$ kV/mm}
\label{t1}
\end{table}

Even at the electric field strength of $5$ kV/mm, the drop in the speed of sound for many fluids is
quite significant. It is quite surprising that despite being a strongly polar fluid the drop in the
speed of sound in polarized water is least among the list. The reason for this is the higher
density of water compared to its relative permittivity. If a fluid like N-methyl formamide can
withstand a field of $8$ kV/mm without suffering a dielectric breakdown then the speed of sound
can drop by $10 \%$ from its value in the absence of electric field. In either cases, the drop in
speed of sound is large enough to be measured experimentally.

\section{Ferrofluids}
Ferrofluids are suspensions of ferromagnetic or ferrimagnetic particles in organic carrier fluids. 
They have no net magnetic moment because the thermal agitations randomly align the individual dipoles
in all directions. However, they give a strong paramagnetic response to an external magnetic field 
\cite{rosensweig2013ferrohydrodynamics}. When an element of a magnetized ferrofluid is compressed
the internal stresses relaxing the compression are opposed by the mutual attraction of the magnetic 
dipoles delaying the process in a manner similar to that in a polarized dielectric fluid. Therefore,
we expect the sound waves to slow down in magnetized ferrofluids as well. The body force density due to 
an external magnetic field is \cite{joshi2013stress}
\begin{equation}\label{e15}
\vec{f} = -\mu_0\frac{H^2}{2}\grad{\kappa_m} + 
\frac{\mu_0}{2}\grad{\left(H^2\rho\pdt{\kappa_m}{\rho}\right)},
\end{equation}
where $H = |\vec{H}|$ is the magnitude of the external magnetizing field, $\kappa_m$ is the relative
permeability of the material and $\mu_0$ is the permeability of free space. The expression for the
body force due to an external magnetizing field is similar to that for the body force due to an
external electric field. We can, therefore, readily conclude that the speed of sound waves in a
ferrofluid is
\begin{equation}\label{e16}
c^2 = c_0^2 - \frac{\mu_0 H^2}{9\rho}(\kappa_m + 2)(\kappa_m - 1)^2
\end{equation}
Following Sommerfeld \cite{sommerfeld2013electrodynamics} we used the Clausius-Mossotti relationship
to get an expression for $\partial\kappa_m/\partial\rho$. For most ferrofluids, the relative 
permeability is less than $5$ at moderately strong magnetizing fields \cite{zakinyan2011drops} making 
the second term in equation \eqref{e16} very small. As a result, we conjecture that the slowing of 
sound speed in a magnetized ferrofluid is extremely difficult to observe in an experiment.

\bibliographystyle{plain}
\bibliography{Slowing_of_sound}
\end{document}